\documentclass[fleqn,usenatbib]{mnras}

\usepackage{newtxtext,newtxmath}

\usepackage[T1]{fontenc}

\DeclareRobustCommand{\VAN}[3]{#2}
\let\VANthebibliography\thebibliography
\def\thebibliography{\DeclareRobustCommand{\VAN}[3]{##3}\VANthebibliography}

\usepackage{multirow}

\usepackage{graphicx}	
\usepackage{amsmath}	
\usepackage{xcolor}     

\usepackage{threeparttable}

\title[Weighing Milky Way Satellites with LISA]{Weighing Milky Way Satellites  with LISA} 

\author[Valeriya Korol et al.]{
Valeriya Korol,$^{1}$\thanks{E-mail: korol@star.sr.bham.ac.uk}
Vasily Belokurov$^{2}$, 
Christopher~J. Moore$^{1}$ 
and Silvia Toonen$^{1}$
\\
$^{1}$Institute for Gravitational Wave Astronomy \& School of Physics and Astronomy, University of Birmingham, Birmingham, B15 2TT, UK \\
$^{2}$Institute of Astronomy, Madingley Rd, Cambridge, CB3 0HA, UK
}

\date{Accepted 2021 January 6. Received 2020 October 8}

\pubyear{2015}

\begin{document}
\label{firstpage}
\pagerange{\pageref{firstpage}--\pageref{lastpage}}
\maketitle

\begin{abstract}

White dwarf stars are a well-established tool for studying Galactic stellar populations. Two white dwarfs in a tight binary system offer us an additional messenger - gravitational waves - for exploring the Milky Way and its immediate surroundings. Gravitational waves produced by double white dwarf (DWD) binaries can be detected by the future {\it Laser Interferometer Space Antenna} (LISA). Numerous and widespread DWDs have the potential to probe shapes, masses and formation histories of the stellar populations in the Galactic neighbourhood. In this work we outline a method for estimating the total stellar mass of Milky Way satellite galaxies based on the number of DWDs detected by LISA. 
To constrain the mass we perform a Bayesian inference using binary population synthesis models and considering the number of detected DWDs associated with the satellite and the measured distance to the satellite as the only inputs.  Based on a fiducial binary population synthesis model we find that for large satellites the stellar masses can be recovered to within 1) a factor two if the star formation history is known and 2) an order of magnitude when marginalising over different star formation history models. For smaller satellites we can place upper limits on their stellar mass. 
Gravitational wave observations can provide mass measurements for large satellites that are comparable, and in some cases more precise, than standard electromagnetic observations.
\end{abstract}

\begin{keywords}
gravitational waves -- binaries: close -- white dwarfs -- Local Group -- galaxies: dwarf
\end{keywords}



\section{Introduction}

White dwarf stars are unique tracers of our Galaxy. Being remnants of old low-mass stars, white dwarfs bear the imprint of the early formation history of the Milky Way inaccessible through more massive stars. The luminosity of a white dwarf depends mainly on its (cooling) age. This property makes white dwarfs a useful tool for dating and reconstructing star formation histories of Galactic stellar populations \citep[e.g.][but see \citealt{tem20}]{tre14,kil17,fan19}.
When two white dwarfs happen to form a short orbital period double white dwarf (DWD) binary they offer us an additional messenger for studying our Galaxy: gravitational waves (GWs).
DWDs are one of the primary targets for the {\it Laser Interferometer Space Antenna} (LISA) designed to detect GWs in the mHz frequency band \citep{LISA}. Tens of thousands of DWDs are expected to be detected by LISA throughout the virial volume of the Galaxy (hundreds of kpc$^3$) and in massive Milky Way satellites \citep[e.g.][]{kor18,kor20,lam19,roe20}. Therefore, they can be used to trace the shape of the Galaxy \citep{ada12,kor19,wil20}. 
We argue that when combined with DWD evolution models, the LISA sample has the potential to probe the Milky Way's properties such as the total stellar mass, the star formation history (SFH) and the initial mass function (IMF). In this work, we make the first steps towards investigating these properties by focusing on the total stellar mass of the Milky Way satellite galaxies.

The stellar mass of a galaxy is often determined from its measured luminosity and an estimate of its stellar-mass-light ratio ($M/L$).
The latter can be derived from stellar population synthesis (SPS) models \citep[e.g.][]{tin76,bru83,ren86, mar98}.
SPS models combine stellar isochrones, spectral libraries and the IMF - each dependent on the metallicity - with a SFH and a dust attenuation law to provide the $M/L$ ratio in a specific electromagnetic band, or even the entire spectral energy distribution of a galaxy.
One can then recover the stellar mass of the galaxy by fitting the obtained SPS models to observations.
Typical uncertainties associated with SPS ingredients such as galaxy evolution and dust prescription result into 0.1-0.3 dex uncertainty on the stellar mass, while the variations in the slope of the IMF may further increase the uncertainty by up to a factor of two \citep{bel01,con09,mcg12}.
Nevertheless, SPS models are one of the key tools for translating between photometry and dynamics.

Building upon the analogy with SPS models, in this work we want to infer the total stellar mass of a galaxy based on the number of detected GW sources. As a proof-of-principle example we focus on the case of Milky Way satellites that requires a very small number of inputs and assumptions.
We anticipate that our method could also be extended for measuring the mass of the Milky Way. However, in addition to resolved DWDs, LISA will also detect an unresolved Galactic stochastic background \citep[e.g.][]{far03,edl05}, which requires a more complex modelling that we defer to a future work. The stochastic background dominates at frequencies <\,3mHz, while DWDs in satellites that LISA can detect are expected to be at higher frequencies.

\section{Satellite Mass inference} \label{sec:model}

A Milky Way satellite with a stellar mass $>10^6\,$M$_\odot$ can host enough DWDs to be identified as a satellite galaxy in the LISA data \citep{roe20}.
We want to estimate the satellite's stellar mass ($M_\star$) given the total number of DWDs detected in the satellite by LISA ($N$) and the satellite's distance ($D$).
The measurement of the distance can come from either electromagnetic and GW observations.

\begin{figure*}
	\includegraphics[width=1.8\columnwidth]{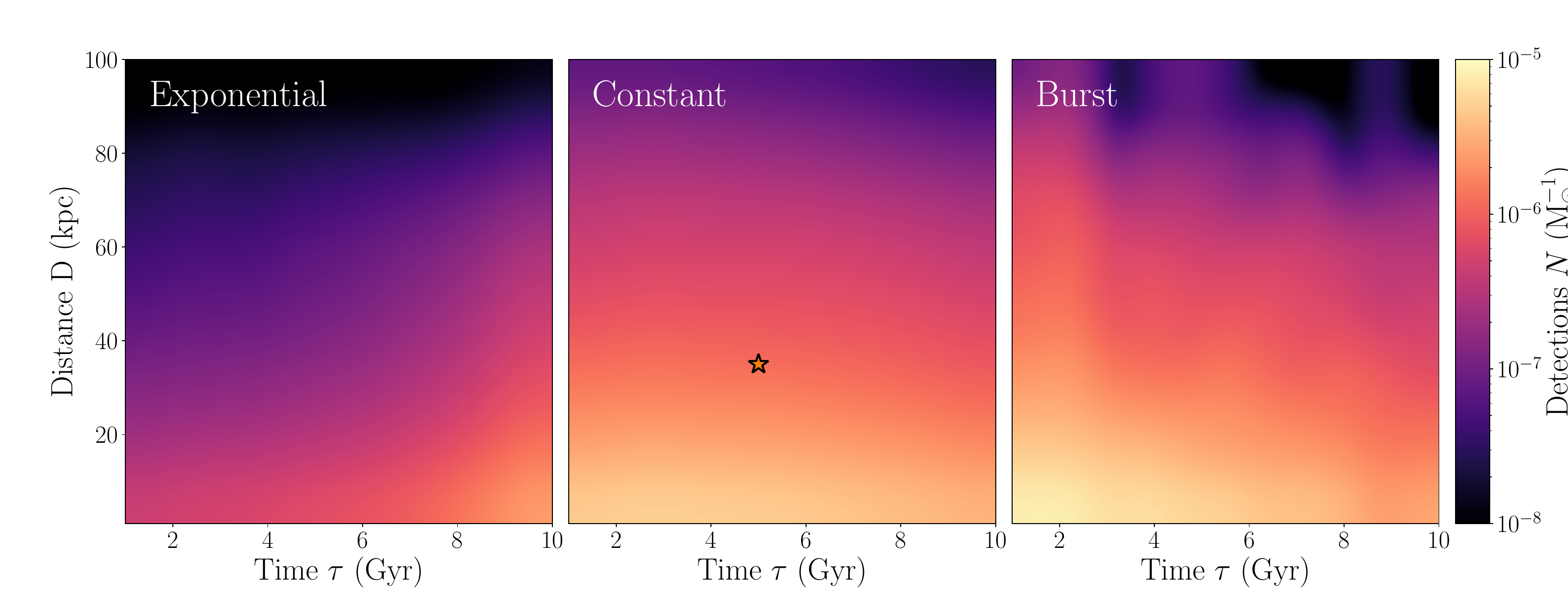}
    \caption{Number of LISA detections per 1\,M$_\odot$ as a function of distance to the satellite and time since the beginning of star formation. Three SFHs are considered (from left to right): exponentially declining with the timescale of 5\,Gyr, constantly star-forming at the rate of 1\,M$_\odot$yr$^{-1}$ and a single burst of star formation $\tau\,$Gyr ago.
    The star represents the test satellite considered in Section~\ref{sec:3.1}.}
    \label{fig:like}
\end{figure*}

Using Bayes' theorem, we write the posterior on the mass as
\begin{equation} \label{eqn:full_post}
    p(M_\star,\tau,D|N) \propto {\cal L}(N|M_\star,\tau,D)\, \pi (M_\star,\tau,D),
\end{equation}
where ${\cal L}(N|M_\star,\tau,D)$ is the likelihood with $\tau$ being  time since the beginning of the star formation, $\pi (M_\star,\tau,D)$ is the prior (we will not need the normalising Bayesian evidence). 
In our inference problem the likelihood is the probability that $N$ sources are produced from a generative model of the satellite galaxy.
Here we adopt satellite models from our previous work based on the DWD evolution models of \citet{too12} with the $\gamma \alpha$ common envelope prescription \citep{nel01}, initial binary fraction of 0.5, the Kroupa IMF \citep{kru93} and metallicity of 0.001 \citep[for further details see][]{kor20}.
These models assume three alternative SFHs: constantly star-forming at 1\,M$_\odot$yr$^{-1}$, exponentially declining with timescale of 5\,Gyr and a single burst.
For a given SFH, our inference problem require the time since the start of the star formation and the distance (in order to assess LISA selection effects) to predict the number of detections at present time per unit Solar mass $\mu(\tau,D)$. 
The number of detections can then be re-scaled to the mass of a satellite; the expected number of sources is written $\lambda=M_\star \mu(\tau,D)$, where $\mu(\tau,D)$ is computed from our models and interpolated over a 2D grid of distances and times (see Fig.~\ref{fig:like}).
In addition, we assume that the LISA event count follows a Poisson distribution, and therefore write
\begin{equation} \label{eqn:like}
     {\cal L}(N|M_\star,\tau,D) = \text{Poisson}\big(N; \lambda[M_\star, \tau,D]\big) = \frac{\lambda^N\exp(-\lambda)}{N!}.
\end{equation}
We chose a uniform prior in $\log M_\star$ (or $\pi[M_\star]\propto1/M_\star$) in the range $10^6$ -- $10^{10}\,$M$_\odot$, where the lower mass end corresponds roughly to a minimum satellite mass required to host at least one LISA detection \citep{kor20} and the upper end to the mass of the largest Milky Way satellite (the Large Magellanic Cloud, LMC).

Because we are interested in the stellar mass, the posterior is marginalised over the distance and time: 
\begin{align} \label{eqn:marg_post}
    p(M_\star|N) &\propto \pi(M_\star) \int \!\mathrm{d}\tau \int\! \mathrm{d}D\; \mathcal{L}\big(N,|M_\star,\tau,D\big) \pi (D) \pi (\tau).
\end{align}
The distance to the satellite is usually known, either from electromagnetic or GW observations. 
Therefore we adopt a Gaussian prior centered on the true distance and with a standard deviation equal to the measurement error.
Unless stated otherwise, we assume a distance error of 10\,per cent, which corresponds roughly to the galaxy radius for a typical satellite in the Galactic halo.
For the constant SFH model we adopt a uniform prior in $\log \tau$ in the range 1 -- 10\,Gyr.
We limit the prior on $\tau$ to 1 -- 6\,Gyrs for the single burst SFH model as it is more appropriate for young stellar populations.
Vice versa, we limit the age prior for the exponential model to 4 -- 10\,Gyr because this SFH is more representative of an old stellar population.

\section{Results} \label{sec:results}

In this section we test the above inference approach in three cases: 1) proof-of-principle example; 2) satellites from a cosmological simulation of a Milky Way-like halo for which we do not know the SFH; and 3) realistic models of known satellites.


{\renewcommand{\arraystretch}{1.5}
\begin{table*}

\centering
\caption{Summary of galaxy properties and the recovered stellar masses from $N$ DWDs detected in a 4 yr LISA mission. The ID indicates the halo number in the simulation suite of \citet{bj05}  with the decimal indicating the sub-halo (satellite galaxy); $\tau$ is taken to be the median age of the simulation-particles constituting the galaxy.}
\label{tab:bj}
\begin{threeparttable}
\begin{tabular}{ccccccccc}
\hline
\multirow{2}{*}{ID} &
  \multirow{2}{*}{$N$} &
  \multirow{2}{*}{$D\,$(kpc)} &
  \multirow{2}{*}{$\tau\,$(Gyr)} &
  \multirow{2}{*}{ $M_\star^{\rm true} (\times 10^6\,$M$_\odot$)} &
  \multicolumn{4}{c}{$M_\star^{\rm est} (\times 10^6\,$M$_\odot$)} \\ \cline{6-9} 
  &    &     &     &      & Constant & Exponential & Burst & Combined \\ \hline 
17.2 & 12 & 53  & 6   & 91   &   $101^{+52}_{-40}$  & $380^{+250}_{-160}$         & $102^{+134}_{-87}$ & $130^{+260}_{-100}$  \\
17.3 & 5  & 35  & 8.4 & 40   & $21.7^{+18.6}_{-8.4}$    &  $103^{+91}_{-61}$        & $61^{+56}_{-28}$  & $27^{+64}_{-19}$ \\ 
7.4 & 2  & 72  & 6.2 & 92   & $24^{+32}_{-16}$    &  $95^{+118}_{-60}$        &  $520_{-480}^{+2170}$ & $ 131^{+316}_{-813}$ \\ 
7.5 & 46  & 85  & 7 & 1266   & $760^{+240}_{-300}$    &  $2460^{+1360}_{-600}$        & $550_{-510}^{+1220}$  & $[100 -1000]$\tnote{*} \\ \hline
\end{tabular}
\begin{tablenotes}
\footnotesize
\item[*] For this case the posterior is multi-modal and we therefore we report the full interval.
\end{tablenotes}
\end{threeparttable}
\end{table*}
}

\subsection{Proof-of-principle} \label{sec:3.1}

Consider a test galaxy with $M_\star^{\rm true} = 4 \times 10^7\,$M$_\odot$, forming stars at a constant rate of 1\,M$_\odot$yr$^{-1}$ for $\tau=5\,$Gyr at a distance of $D = 35\,$kpc. For these parameters our fiducial DWD population synthesis model predicts 9 LISA detections (cf. star in Fig.~\ref{fig:like}).  We now use the inference method described in Section~\ref{sec:model} to show that we can recover the original stellar mass.

We sample the posterior (Eq.~\ref{eqn:full_post}) using the affine invariant sampler \texttt{emcee} \citep{emcee} before marginalising (Eq.~\ref{eqn:marg_post}) to measure the stellar mass.
We recover $M_\star^{\rm est} = 4.2_{-1.5}^{+2.1} \times 10^7\,$M$_\odot$, where the errors represent $1\sigma$ uncertainty. 
The uncertainty on the mass depends only weakly on the distance measurement; reducing (increasing) the uncertainty to $5\%$ ($30\%$) changes the mass estimate to $4.1_{-1.4}^{+2.1} \times 10^7\,$M$_\odot$ ($4.2_{-2.1}^{+3.5} \times 10^7\,$M$_\odot$).
Instead it mostly comes from the uncertainty on the age and the Poisson fluctuations on the DWD count;
indeed, when applying a more stringent prior on the satellite's age (a Gaussian centered on the true age $\pm$1\,Gyr, we reduce the uncertainty on the mass by 50\, per cent). 

We now examine the sensitivity of our mass measurement to the assumed SFH. If we re-analyse the same example `incorrectly' using the exponential SFH model we obtain $M_\star^{\rm est} = 14.8_{-6.0}^{+9.3
} \times 10^7\,$M$_\odot$, which over-estimates the satellite's mass by $\sim1.8\sigma$.
The single burst model yields a more precise measurement of $M_\star^{\rm est} = 6.3_{-2.9}^{+5.6} \times 10^7\,$M$_\odot$.
This behaviour can be understood with the aid of Fig.~\ref{fig:N_vs_age} which shows the number of LISA detections for the test satellite predicted by the three SFH models as a function of time.
The constant SFH model is consistent with the observed number of sources (Poissonian error $9 \pm \sqrt{9}$) for all times.
The exponential model predicts fewer sources leading to an overestimate of the mass as a compensation.
Because no new binaries are formed at $\tau>0$, the single burst SFH model is shaped by binary evolution with the peaks corresponding to the typical timescales required for DWDs of different masses to enter the LISA band (e.g. the first peak rises from more massive but short-lived carbon-oxygen white dwarfs binaries; cf. Fig.~1 of \citealt{kor20}). It remains below the constant SFH model for $\tau>1$\,Gyr and overlaps with the source count uncertainty for $\tau < 6\,$Gyr, meaning that the single burst model will tend to over-estimate the mass compared to the true value. 
An example of posteriors on the free parameters of our inference problem ($\tau,\, D$ and $M_\star$) is shown in Fig.~\ref{fig:posteriors_test}. 
It demonstrates that counting the number of LISA sources in a satellite does not constrain the age (i.e. time science star formation) or distance (as we essentially recover the priors), but does provide a measurement of the stellar mass.
Finally, if the SFH is uncertain, it is possible to marginalise over different scenarios.
Marginalising over the three considered SFHs (with equal prior weights) for our test galaxy gives $M_\star^{\rm est} = 4.8_{-2.9}^{+9.6} \times 10^7\,$M$_\odot$.

Finally, we want to assess the robustness of our inference model against assumptions on the DWD population such as the initial binary fraction, metallicity and IMF. 
In re-scaling the number of detected DWDs to assume the initial binary fraction of 30 -- 90 per cent, we find the recovered mass using the fiducial values stays within $1\sigma$ from the assumed ‘true’ mass.
In addition, we test IMFs of \citet{mil79} and \citet{sca86}. Although both favour more DWD progenitors compared to the Kroupa IMF, we find a similar number of LISA detections as for the fiducial model \citep[for discussion see Sect. 4.2 of][]{kor20}. For metallicity values of 0.0001 and 0.02 we obtain 7 and 9 detectable DWDs for the test satellite.  Thus, variations in IMF and metallicity does not significantly influence the accuracy of our mass estimates.
The largest variation in the number of detectable DWDs come from the assumptions on the common envelope evolution \citep{too17}; in \citet{kor20} the largest variation among the satellite models is less than a factor of 2.
Because the posterior of the satellite mass is not symmetric, by increasing the number of detections by a factor of 2 we overestimate the mass by 2$\sigma$, while by decreasing the number of detections our mass estimate stays within 1$\sigma$ for all SFHs.

\subsection{Synthetic satellites with different SFHs} \label{sec:syn_sats}

Now we would like to test our method on a few cases with the SFHs that differ from our models described above. In the absence of the real data we construct a mock dataset by combining our DWDs evolution models with cosmological simulations of the Milky Way-like halos from \citet{bj05}. Specifically, for this work we employed their halo 17 and halo 07. 
We assign a number of DWDs to particles in  each simulation based on the particle's mass and randomly draw binaries from our fiducial DWD evolution model with $\gamma\alpha-$common envelope prescription \citep{nel01}, metallicity of 0.001, the IMF of \citet{kru93}, and initial binary fraction of 0.5 \citep[for details see][]{too12}. 
We then model the orbital evolution of the binaries due to the GW radiation  starting from the DWD formation until the age of the simulation particle. We discard binaries if their formation times are greater than the age of the simulation particle -- they have begun mass transfer (i.e. when one of the white dwarfs fills its Roche lobe) or they have already merged within this time.
We compute the detectability of the DWDs with LISA as described in Section 2.4 of \citet{kor20}. 
The detection threshold for a signal-to-noise of 7 after 4\,years of observations with LISA yields DWD detections from a number of satellites with masses of $10^7 - 10^{9}\,$M$_\odot$.

\begin{figure} 
	\includegraphics[width=0.8\columnwidth]{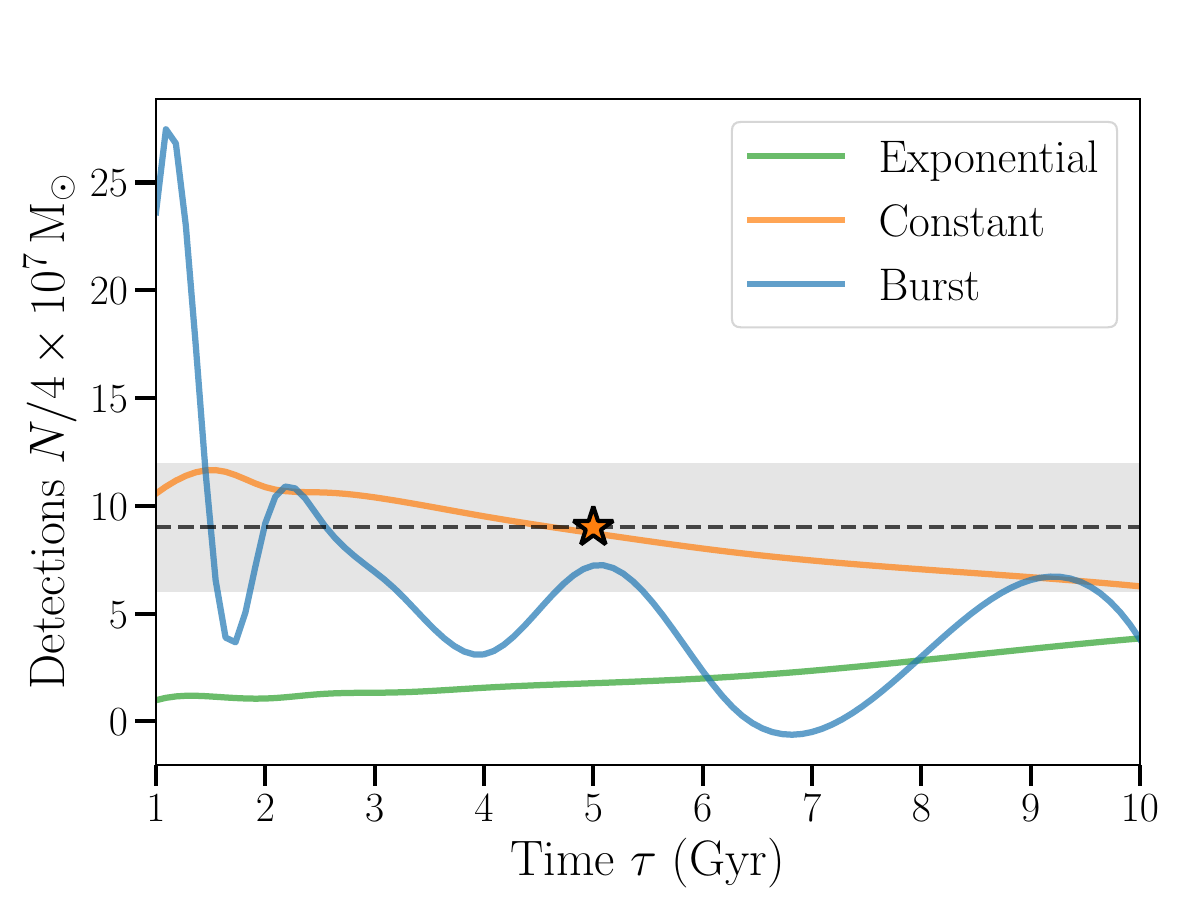}
    \caption{Number of LISA detections as a function of time for the test since the beginning of star formation for the test galaxy with $M_\star = 4 \times 10^7\,$M$_\odot$ and $D=35\,$kpc.
    The SFH models are an exponential (green), constant (orange) and single burst (blue). The star represents the number of detections corresponding for the test satellite at $\tau=5$\,Gyr. 
    The dashed horizontal line represents the number of detections in the test example while shaded region represents the respective Poissonian error ($9 \pm \sqrt{9}$)
    }
    \label{fig:N_vs_age}
\end{figure}

\begin{figure} 
	\includegraphics[width=1\columnwidth]{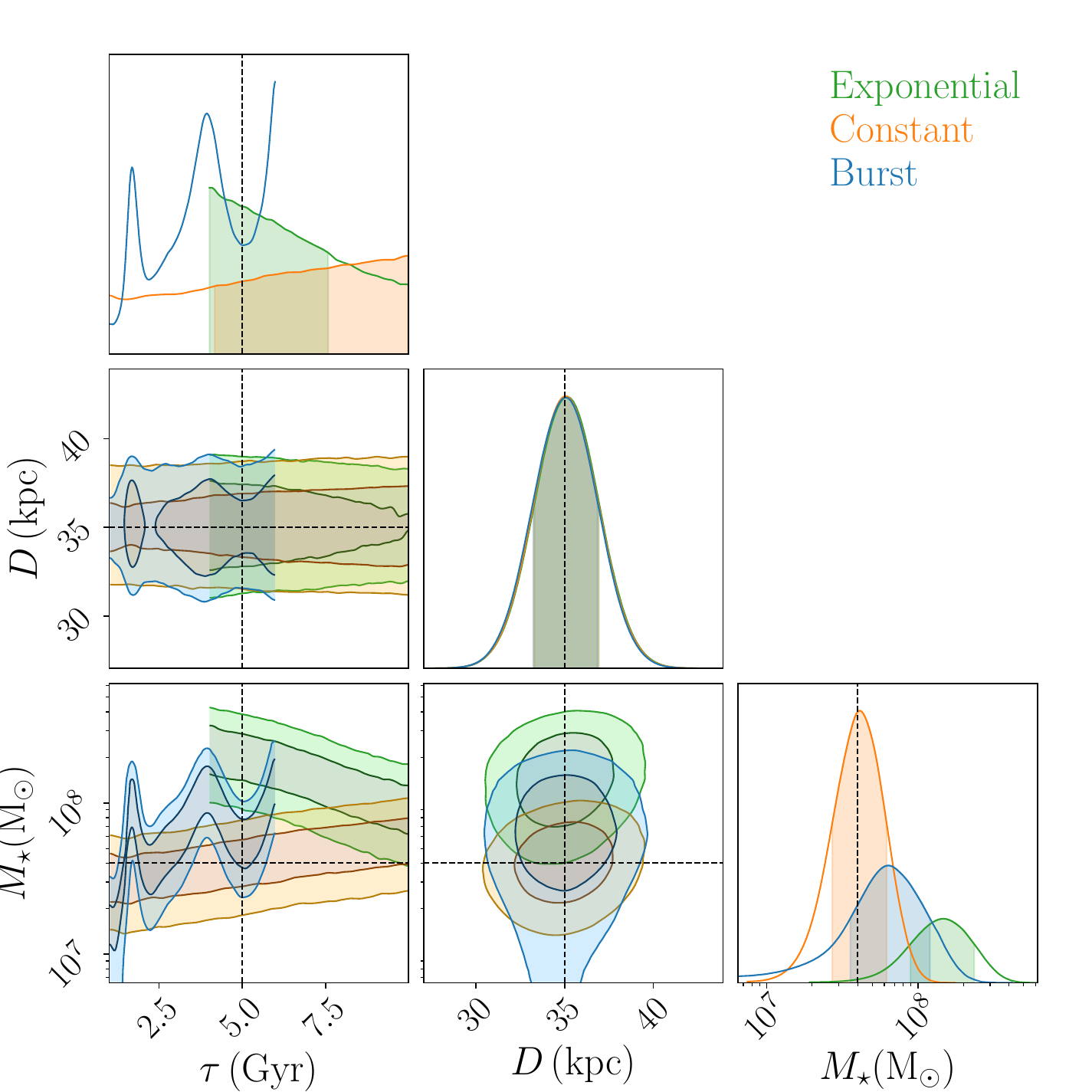}
    \caption{Posteriors for the test galaxy. 
    From this corner plot it can be seen that counting the number of LISA sources in a satellite does not provide a good measurement of the age. It also does not provide a good measurement of the distance (we essentially recover the Gaussian prior). However, it does provide a measurement of the stellar mass.
    }
    \label{fig:posteriors_test}
\end{figure}

We list the true properties and the recovered total stellar masses of the mock satellites in Table~\ref{tab:bj}.
We note that overall, the constant SFH model is more precise compared to the other two SFH models, but it starts to be biased towards lower satelite masses when the number of detections is too few.
The exponential SFH model works better when the number of detections is low; it yields a larger confidence interval.
The model with the burst SFH tends to give the largest (fractional) uncertainty on the stellar mass. 
This can be traced back to the oscillatory behaviour in Fig.~\ref{fig:N_vs_age}. 
The expected number of sources oscillates and we tend to recover multi-modal posteriors on $\tau$ with each mode giving a slightly different mass estimate; marginalising over $\tau$ then tends to give a larger final uncertainty.

When marginalising over all possible SFHs (final column of Table~\ref{tab:bj}) we find very large uncertainties in some cases. 
We stress however, that this is a very conservative and unrealistic case that assumes almost no knowledge of the history of the satellite.
In order to measure the mass of a satellite galaxy in this way requires some knowledge of the SFH; this is analogous to the way in which SPS models rely on prior knowledge of the physics of stellar isochrones, the galaxy IMF etc.

\subsection{Known satellites} \label{sec:known_sats}

Next, we would like to estimate the precision of our method for some of the known MW satellites. To do this we use predictions for the number of detectable binaries based on realistic SFHs for Sagittarius and LMC. We recover their stellar mass using the constant SFH model.

Using a multi-burst SFH, \citet{kor20} forecast 10 detectable DWDs in the Sagittarius galaxy adopting the `true' (current) mass of the remnant of $2.1 \times 10^7$M$_\odot$  \citep{McC12,VB2020}.
We obtain a mass of $3.1_{-1.0}^{+1.5} \times 10^7$M$_\odot$, consistent with the true mass.
Given that the original stellar mass of the Sagittarius galaxy is still uncertain and can be as high as $5.5 \times 10^8\,$M$_\odot$ \citep[see][]{NO2010,joh20}, a new measurement based on GW detections will provide an important independent constraint. 

\citet{kei20} constructed a realistic DWD catalogue for the LMC using a SFH derived from the spatially resolved color-magnitude diagrams by \citet{har09} and spatial stellar mass distribution from the numerical simulation of \citet{luc19}. Assuming $M_{\star {\rm LMC}}^{\rm true} = 2.7 \times 10^9\,$M$_\odot$ \citep{van20} they obtained 333 detached DWDs detectable by LISA for the nominal 4\,yr of mission. We input this number into our inference machinery adopting the constant SFH model and a Gaussian prior on the distance centered on 49.97\,kpc \citep{pie13} with a standard deviation of 5\,kpc (corresponding roughly to the radius of the LMC). We recover $M_{\star {\rm LMC}}^{\rm est} =2.4^{+0.6}_{-0.8}\times 10^9$ consistent with the true value within $1\sigma$. Our uncertainty on the LMC stellar mass is comparable with that reported in \citet{van20} derived by assuming $M/L_V = 0.9 \pm 0.2$ \citep{bel01} with $L_V= 30 \times 10^9\,$L$_\odot$. 

Finally, when considering numerous smaller satellites it is likely that LISA will detect no DWDs. In this case our approach allows us to derive an upper limit for the satellite's stellar mass. For example, \citet{kor20} find that Fornax and Sculptor galaxies will unlikely host detectable LISA sources because of their low masses ($2.0 \times 10^7$ and $2.6 \times 10^6\,$M$_\odot$ respectively) and/or quite large distances (147 and 86\,kpc) \citep{McC12}. By running our inference model with $N=0$ detections we can say that $M_\star< 6.2 \times 10^7\,$M$_\odot$ for Fornax and $M_\star <4.3 \times 10^7\,$M$_\odot$ for Sculptor - with 90 per cent confidence.

\section{Discussion and Conclusions} \label{sec:last}

In this work we demonstrated how the total stellar mass of the Milky Way satellites can be estimated from the number of associated LISA GW events. 
Using a fiducial DWD evolution model we showed that satellite masses inside the Milky Way virial radius can be recovered within 1) a factor two if the SFH is known and 2) within an order of magnitude even when marginalising over alternative SFHs. 
This later case is rather pessimistic because SFHs of the local dwarf satellites are extensively studied in the literature \citep[e.g.][for a review]{wei14}.
Any SFH, once known, can be implemented in our model as was done here for three representative examples (exponential, constant and single burst); these examples can be further combined to obtain more complex SFHs.
We find that the constant SFH - being less dependent on the satellite's mean age (see Fig.~\ref{fig:like} and \ref{fig:N_vs_age}) - yields smaller uncertainties compare to the exponential and a single burst SFHs, and works well across several examples where the number of GW events is more than a few (cf. Sec.~\ref{sec:syn_sats}). 

The accuracy of our method mainly depends on the prior knowledge of the SFH and on the number of LISA detections, increasing with the number of detections (Poisson errors scaling as $1/\sqrt{N}$.) 
If no GW events are identified (e.g. when $M_\star < 10^6\,$M$_\odot$ and/or $D >100\,$kpc and/or the Galactic foreground is comparable to $N$ at the satellite's position) our method would still allow to define an upper limit of the satellite's mass. 
Other assumptions such as uncertainties on the distance and the age of the satellite have smaller effects on the results.  
We did not quantify the effect of uncertainties in the DWD evolution modelling. Throughout this work we have assumed a fiducial model of \citet[][based on the $\gamma\alpha$-common envelope evolution]{too12} that yields the space density of DWDs in agreement with observations of the local white dwarf population \citep{max99,too17}. Within this model, we found that the variations of the binary fraction, IMF and metallicity do not significantly influence the accuracy of the mass estimate.
Other recent models reported space densities a factor of two lower compared to our fiducial model \citep[e.g.][]{lam19,bre20}.
However, in the next decades the differences between binary evolution models should decrease as more observations will become available for calibration \citep[e.g. space densities of the shortest period DWDs detectable as eclipsing binaries,][]{kor17}.

The concept of the method presented here is similar to that using $M/L$ relations from SPS models. 
Although our method is still affected by some of the similar systematic uncertainties, it does not require an assumption on the dust attenuation and, as has been demonstrated, is relatively mildly affected by some of assumptions employed in DWD binary evolution models (cf. Sec.~\ref{sec:3.1}).
Note however that alternative common envelop prescriptions may lead to differences of a factor of two \citep{too17,kor20}. 
In contract to the $M/L$ based estimates that are sensitive to the mass enclosed in bright stars, GW detections yield the original stellar mass of a satellite including the contribution due to evolved stars that are no longer visible through light.

Our method can be refined in many ways. For example, as inputs in the inference model, we considered only the number of DWDs and the distance. GW observations will also provide frequency and chirp mass distributions of DWDs that are indicative of the satellite's SFH \citep{kei20}, and therefore can be implemented in the model.
We focused on DWDs exclusively, but our method can be extended to other double compact objects. Black hole and neutron star binaries are stronger GW emitters compared to DWDs,  but only a few are expected to be detected in the most massive satellites like the Magellanic Clouds \citep[e.g.][]{lau19}.  Thus, extending our model to other double compact objects binaries could improve the predicted number counts and the distance horizon for applicability of the method, but would not significantly influence the estimate of the mass. However, being more sensitive to the metallicity and star formation rate, binary black holes and neutron stars have been proposed to trace these properties of their host galaxies across cosmic time \citep[e.g.][]{art19,chr19}.

Our inference method can be extended to measuring the mass of the Milky Way.
This would require constructing a more complex hierarchical Bayesian model accounting also for the spatial distribution of the LISA detections \citep[e.g.][]{ada12}. In addition, it is necessary to account for the fact that the majority of the DWDs in the Milky Way will be unresolved and will form a confusion background. The shape of the confusion background would provide additional information that can be fed into the model \citep[e.g.][]{ben06,bre19}. 
 
Based on electromagnetic studies, the stellar masses of the Milky Way satellites - even of the most studied ones like the LMC- are still relatively uncertain, with typical uncertainties within a factor of a few \citep{bel01,mcg12}. 
With the addition of independent measurements based on GW observations, we can place stronger constraints on their masses. Furthermore, we can gain a deeper understanding of the other satellite's properties and their co-evolution with the Milky Way.

\section*{Acknowledgements}
We thank Michael Keim for providing us some of the results of his master thesis.
We also thank Riccardo Buscicchio, Davide Gerosa, Antoine Klein, Luke T. Maud, Elena Sacchi, Elinore Roebber and Alberto Vecchio for helpful discussions and suggestions.
VK and ST acknowledge support from the Netherlands Research Council NWO (respectively Rubicon 019.183EN.015 and VENI 639.041.645 grants). 
This work made use of the suite of Milky Way-like halo simulations of \citet{bj05} publicly available at \url{http://user.astro.columbia.edu/~kvj/halos/},
the \texttt{emcee} MCMC \texttt{python} sampling toolkit \citep{emcee} and \texttt{ChainConsumer python} plotting package \citep{chainconsumer}.

\section*{Data Availability}

The data underlying this article and our code are publicly available at \url{https://github.com/korolvaleriya/Weighing-Milky-Way-Satellites-with-LISA}


\bibliographystyle{mnras}
\bibliography{biblio}

\bsp	
\label{lastpage}
\end{document}